\documentclass[runningheads]{llncs}
\usepackage{caption}
\usepackage{subcaption}
\usepackage{booktabs}
\usepackage{amsmath}
\usepackage{amsfonts}
\usepackage{multirow}
\usepackage{lipsum}

\usepackage[T1]{fontenc}
\usepackage{graphicx}

\newcommand{\figref}[1]{\figurename~\ref{#1}}
\newcommand{\tabref}[1]{\tablename~\ref{#1}}
\newcommand{\secref}[1]{Section~\ref{#1}}

\usepackage{hyperref}
\usepackage{xcolor}

\begin{document}
\title{Deep Homography Prediction for Endoscopic Camera Motion Imitation Learning}
%
%

\author{Martin Huber\inst{1}\orcidID{0000-0003-4603-6773} \and S\'{e}bastien Ourselin\inst{1}\orcidID{0000-0002-5694-5340} \and
Christos Bergeles\inst{1}\thanks{These authors contributed equally to this work}\orcidID{0000-0002-9152-3194} \and Tom Vercauteren\inst{1}$^\star$\orcidID{0000-0003-1794-0456}}


\authorrunning{M. Huber et al.}
\titlerunning{Deep Homography Prediction for Camera Motion Learning}

%
\institute{King's College London, School of Biomedical Engineering \& Image Sciences, United Kingdom}

\maketitle              
\begin{abstract}In this work, we investigate laparoscopic camera motion automation through imitation learning from retrospective
videos of laparoscopic interventions. A novel method is introduced that learns to augment a surgeon's behavior in image space through object motion invariant image registration via homographies. Contrary to existing approaches, no geometric assumptions are made and no depth information is necessary, enabling immediate translation to a robotic setup.
Deviating from the dominant 
approach in the literature which consist of following a surgical tool, we do not handcraft the objective and no priors are imposed on the surgical scene, allowing the method to discover unbiased policies. In this new research field, significant improvements are demonstrated over two baselines on the Cholec80 and HeiChole datasets, showcasing an improvement of $47\%$ over camera motion continuation. The method is further shown to indeed predict camera motion correctly on the public motion classification labels of the AutoLaparo dataset. All code is made accessible on GitHub\footnote{\href{https://github.com/RViMLab/homography\_imitation\_learning}{https://github.com/RViMLab/homography\_imitation\_learning}}.

\end{abstract}

\section{Introduction}
Automation in robot-assisted minimally invasive surgery (RMIS) may reduce human error that is linked to fatigue, lack of attention and cognitive overload \cite{fiorini2022concepts}.
It could help surgeons operate such systems by reducing the learning curve \cite{van2018learning}.
And in an ageing society with shrinking workforce, it could help to retain accessibility to healthcare. It is therefore expected that parts of RMIS will be ultimately automated \cite{davenport2019potential,zidane2022robotics}. On the continuous transition towards different levels of autonomy, camera motion automation is likely to happen first \cite{kitaguchi2022artificial}.

Initial attempts to automate camera motion in RMIS include rule-based approaches that keep surgical tools in the center of the field of view \cite{da2020scan,garcia2022robotic,sandoval2021towards}. The assumption that surgical tools remain centrally is, however, simplistic, as in many cases the surgeon may want to observe the surrounding anatomy to decide their next course of action. 

Contrary to rule-based approaches, data-driven methods are capable to capture more complex control 
policies.
Example data-driven methods suitable for camera motion automation include reinforcement learning (RL) and imitation learning (IL). The sample inefficiency and potential harm to the patient currently restrict RL approaches to simulation \cite{su2021multicamera,scheikl2023lapgym,agrawal2018automating}, where a domain gap remains. Work to bridge the domain gap and make RL algorithms deployable in real setups have been proposed \cite{cartucho2021visionblender,marzullo2021towards}, but clinical translation has not yet been achieved. For IL, on the other hand, camera motion automation could be learned from real data, thereby implicitly tackling the domain-gap challenge. The downside is that sufficient data may be difficult to collect. Many works highlight that lack of expert annotated data hinders progress towards camera motion automation in RMIS \cite{maier2022surgical,kassahun2016surgical,esteva2019guide}.
It is thus not surprising that existing literature on IL for camera motion automation utilizes data from mock setups \cite{ji2018learning,wagner2021learning}.

Recent efforts to make vast amounts of laparoscopic intervention videos publicly available \cite{maier2022surgical} drastically change how IL for camera motion automation can be approached. So far, this data is leveraged mainly to solve auxiliary tasks that could contribute to camera motion automation. As reviewed in \cite{loukas2018video}, these tasks include tool and organ segmentation, as well as surgical phase recognition. For camera motion automation specifically, however, there exist no publicly available image-action pairs. Some work, therefore, continues to focus on the tools to infer camera motion \cite{li2021data}, or learns on a robotic setup altogether \cite{li20223d} where camera motion is accessible. The realization, however, that camera motion is intrinsic to the videos of laparoscopic interventions and that camera motion could be learned on harvested actions was first realized in \cite{huber2022deep}, and later in \cite{li2022learning}. This comes with the additional advantage that no robot is necessary to learn behaviors and that one can directly learn from human demonstrations.

In this work, we build on \cite{huber2022deep} for computationally efficient image-action pair extraction from publicly available datasets of laparoscopic interventions, which yields more than $20\times$ the amount of data that was used in the closed source data of \cite{li2022learning}. Contrary to \cite{li2022learning}, our camera motion extraction does not rely on image features, which are sparse in surgical videos, and is intrinsically capable to differentiate between camera and object motion. We further propose a novel importance sampling and data augmentation step for achieving camera motion automation IL.









\section{Materials and Methods}
\begin{figure}
    \centering
    \includegraphics[scale=.6]{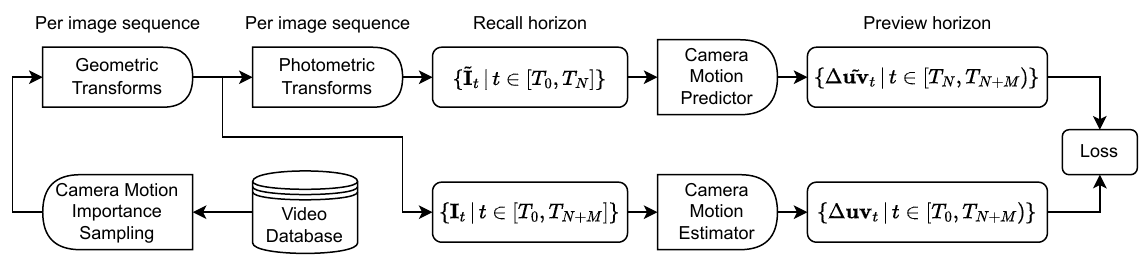}
    \caption{Training pipeline, refer to  \secref{sec:proposed_pipeline}. From left to right: Image sequences are importance sampled from the video database and random augmentations are applied per sequence online. The lower branch estimates camera motion between subsequent frames, which is taken as pseudo-ground-truth for the upper branch, which learns to predict camera motion on a preview horizon.}
    \label{fig:training_pipeline}
\end{figure}

The proposed approach to learning camera motion prediction is summarized in \figref{fig:training_pipeline}. The following sections will describe its key components in more detail.

\subsection{Theoretical Background}
\label{sec:theoretical_background}
Points on a plane, as observed from a moving camera, transform by means of the $3\times3$ projective homography matrix $\mathbf{G}$ in image space. Thus, predicting future camera motion (up to scale) may be equivalently treated as predicting future projective homographies.

It has been shown in \cite{DeTone2016DeepEstimation} that the four point representation of the projective homography, \emph{i.e.,}  taking the difference between four points in homogeneous coordinates $\Delta\mathbf{uv} = \{\mathbf{p}_i - \mathbf{p}^\prime_i\,|\,i \in [0, 4)\} \in \mathbb{R}^{4\times2}$ that are related by $\mathbf{G}\mathbf{p}_i \sim \mathbf{p}^\prime_i\,\,\forall i$, is better behaved for deep learning applications than the $3\times3$ matrix representation of a homography.
Therefore, in this work, we treat camera motion $\mathcal{C}$ as a sequence of four point homographies on a time horizon $[T_0, T_{N+M})$, $N$ being the recall horizon's length, $M$ being the preview horizon's length. Time points 
lie $\Delta t$ apart, that is $T_{i+1} = T_{i} + \Delta t$. For image sequences of length $\text{N+M}$,
we work with
four point homography sequences $\mathcal{C}=\{\Delta\mathbf{uv}_t \,|\, t\in[T_0, T_{N+M})\}$.

\subsection{Data and Data Preparation}
\label{sec:data_and_data_preparation}
Three datasets are
curated
to train and evaluate the proposed method: two cholecystectomy datasets (laparoscopic gallbladder removal), namely Cholec80 \cite{twinanda2016endonet} and HeiChole \cite{wagner2023comparative}, and one hysterectomy dataset (laparoscopic uterus removal), namely AutoLaparo \cite{wang2022autolaparo}.


To remove status indicator overlays from the laparoscopic videos, which may hinder the camera motion estimator, we identify the bounding circle of the circular field of view using \cite{Budd2022RapidDataset}. We crop the view about the center point of the bounding circle to a shape of $240\times320$, so that no black regions are prominent in the images.


All three datasets are split into training, validation, and testing datasets. We split the videos by frame count into $80\pm 1\,\%$ training and $20 \pm 1\,\%$ testing. Training and testing videos never intersect. We repeat this step to further split the training dataset into (pure) training and validation datasets.

Due to errors during processing the raw data, we exclude videos $19$, $21$, and $23$ from HeiChole, as well as videos $22$, $40$, $65$, and $80$ from Cholec80. This results in dataset sizes of: Cholec80 - $4.4e6$ frames at $25\,\text{fps}$, HeiChole - $9.5e5$ frames at $25\,\text{fps}$, and AutoLaparo - $7.1e4$ frames at $25\,\text{fps}$.


\subsection{Proposed Pipeline}
\label{sec:proposed_pipeline}


\subsubsection{Video Database and Importance Sampling}
\label{sec:video_database_and_importance_sampling}
The curated data from \secref{sec:data_and_data_preparation} is accumulated into a video database. Image sequences of length $N+M$ are sampled at a frame increment of $\Delta n$ between subsequent frames and with $\Delta c$ frames between the sequence's initial frames. Prior to adding the videos to the database, an initial offline run is performed to estimate camera motion $\Delta\mathbf{uv}$ between the frames. This creates image-motion correspondences of the form $(\mathbf{I}_n\,,\mathbf{I}_{n+\Delta n}\,,\Delta\mathbf{uv}_n)$. Image-motion correspondences where $\mathbb{E}(||\Delta\mathbf{uv}_n||_2) > \sigma$, with sigma being the standard deviation over all motions in the respective dataset, define anchor indices $n$. Image sequences are sampled such that the last image in the recall horizon lies at index $n = N-1$, marking the start of a motion. The importance sampling samples indices from the intersection of all anchor indices, shifted by $-N$, with all possible starting indices for image sequences.

\subsubsection{Geometric and Photometric Transforms}
\label{sec:geometric_and_photometric_transforms}
The importance sampled image sequences are fed to a data augmentation stage. This stage entails geometric and photometric transforms. The distinction is made because downstream, the pipeline is split into two branches. The upper branch serves as camera motion prediction whereas the lower branch serves as camera motion estimation, also refer to the next section.
As it acts as the source of pseudo-ground-truth,
it is crucial that the camera motion estimator performs under optimal conditions, hence no photometric transforms, i.e. transforms that change brightness / contrast / fog etc., are applied. Photometrically transformed images shall further be denoted as $\tilde{\mathbf{I}}$. To encourage same behavior under different perspectives, geometric transforms are applied, i.e. transforms that change orientation / up to down / left to right etc. Transforms are always
sampled randomly, and 
applied consistently to the entire image sequence.
 
\subsubsection{Camera Motion Estimator and Predictor}
\label{sec:camera_motion_estimator_and_predictor}
The goal of this work is to have a predictor learn camera motion computed by an estimator.
The predictor takes as input a photometrically and geometrically transformed recall horizon $\{\tilde{\mathbf{I}}_t\,|\,t\in[T_0, T_N)\}$ of length $N$, and predicts camera motion $\tilde{\mathcal{C}} = \{\Delta \tilde{\mathbf{uv}}_t\,|\,t\in[T_N,T_{N+M})\}$ on the preview horizon of length $M$. The estimator takes as input the geometrically transformed preview horizon $\{\mathbf{I}_t\,|\,t\in[T_M, T_{N+M})\}$ and estimates camera motion $\mathcal{C}$, which serves as a target to the predictor. The estimator is part of the pipeline to facilitate on-the-fly perspective augmentation via the geometric transforms.




\section{Experiments and Evaluation Methodology}
The following two sections elaborate the experiments we conduct to investigate the proposed pipeline from \figref{fig:training_pipeline} in \secref{sec:proposed_pipeline}. First the camera motion estimator is investigated, followed by the camera motion predictor.

\subsection{Camera Motion Estimator}
\label{sec:camera_motion_estimator}
\subsubsection{Camera Motion Distribution} 
To extract the camera motion distribution, we run the camera motion estimator from \cite{huber2022deep} with a ResNet-34 backbone over all datasets from \secref{sec:data_and_data_preparation}. We map the estimated four point homographies to up/down/left/right/zoom-in/zoom-out for interpretability. Left/right/up/down corresponds to all four point displacements $\Delta\mathbf{uv}$ consistently pointing left/right/{\allowbreak}up/down respectively.
Zoom-in/out corresponds to all four point displacements $\Delta\mathbf{uv}$ consistently pointing inwards/outwards. Rotation left corresponds to all four point displacements pointing up right, bottom right, and so on. Same for rotation right. Camera motion is defined static if it lies below the standard deviation in the dataset. The frame increment is set to $0.25\,\text{s}$, corresponding to $\Delta n = 5$ for the $25\,\text{fps}$ videos.

\subsubsection{Online Camera Motion Estimation}
\label{sec:online_camera_motion_estimation}
Since the camera motion estimator is executed online, memory footprint and computational efficiency are of importance. Therefore, we evaluate the estimator from \cite{huber2022deep} with a ResNet-34 backbone, SURF \& RANSAC, and LoFTR \cite{sun2021loftr} \& RANSAC. Each estimator is run 1000 times on a single image sequence of length $N+M=15$ with an NVIDIA GeForce RTX 2070 GPU and an Intel(R) Core(TM) i7-9750H CPU @ 2.60GHz.

\subsection{Camera Motion Predictor}
\label{sec:camera_motion_predictor_experiments}
\subsubsection{Model Architecture}
For all experiments, the camera motion predictor is a ResNet-18/34/50, with the number of input features equal to the recall horizon $\text{N}\times3$ (RGB), where $\text{N}=14$. We set the preview horizon $\text{M}=1$. The frame increment is set to $0.25\,\text{s}$, or $\Delta n = 5$ for the $25\,\text{fps}$ videos. The number of frames between clips is also set to $0.25\,\text{s}$, or $\Delta c = 5$.

\subsubsection{Training Details}
The camera motion predictor is trained on each dataset from \secref{sec:data_and_data_preparation} individually. For training on Cholec80/HeiChole/AutoLaparo, we run $80/50/50$ epochs on a batch size of $64$ with a learning rate of $2.5e-5/1.e-4/1.e-4$. The learning rates for Cholec80 and HeiChole relate approximately to the dataset's training sizes, see \ref{tab:camera_motion_prediction}. For Cholec80, we reduce the learning rate by a factor $0.5$ at epochs $50,\,75$. For Heichole/AutoLaparo we drop the learning rate by a factor $0.5$ at epoch $35$. The loss in \figref{fig:training_pipeline} is set to the mean pairwise distance between estimation and prediction $\mathbb{E}(||\Delta\tilde{\mathbf{uv}}_t - \Delta\mathbf{uv}_t||_2) + \lambda \mathbb{E}(||\Delta\tilde{\mathbf{uv}}_t||_2)$ with a regularizer that discourages the identity $\Delta\tilde{\mathbf{uv}}_t = \mathbf{0}$ (i.e. no motion). We set $\lambda = 0.1$.

\subsubsection{Evaluation Metrics}
For evaluation we compute the mean pairwise distance between estimated and predicted motion $\mathbb{E}(||\Delta\tilde{\mathbf{uv}}_t - \Delta\mathbf{uv}_t||_2)$. All camera motion predictors are benchmarked against a baseline, that is a $\mathcal{O}(1)$/$\mathcal{O}(2)$-Taylor expansion of the estimated camera motion $\Delta\mathbf{uv}_t$. Furthermore, the model that is found to perform best is evaluated on the multi-class labels (left, right, up, down) that are provided in AutoLaparo. 

\section{Results}
\subsection{Camera Motion Estimator}
\subsubsection{Camera Motion Distribution}
\begin{figure}
    \centering
    \begin{subfigure}[b]{0.49\textwidth}
        \centering
        \includegraphics[width=\textwidth]{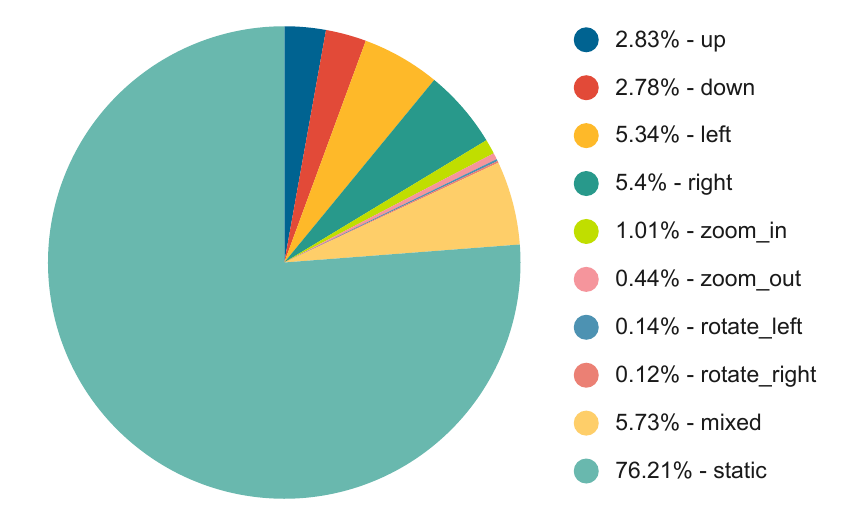}
        \caption{Cholec80}
    \end{subfigure}
    \hfill
    \centering
    \begin{subfigure}[b]{0.49\textwidth}
        \centering
        \includegraphics[width=\textwidth]{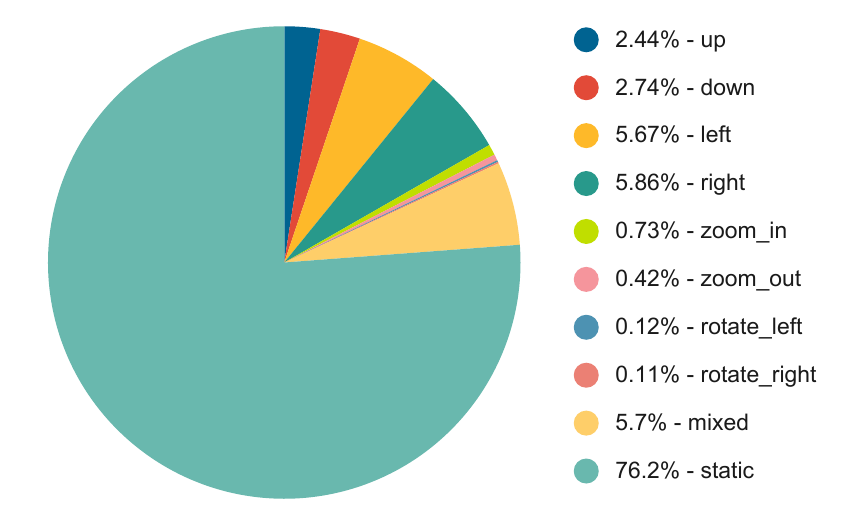}
        \caption{HeiChole}
    \end{subfigure}
    \caption{Camera motion distribution, refer to \secref{sec:camera_motion_estimator}. AutoLaparo: $2.81\%$ - up, $1.88\%$ - down, $4.48\%$ - left, $3.38\%$ - right, $0.45\%$ - zoom\_in, $0.2\%$ - zoom\_out, $0.3\%$ - rotate\_left $0.3\%$, - rotate\_right $14.9\%$ - mixed, $71.29\%$ - static.}
    \label{fig:camera_motion_distribution}
\end{figure}

The camera motion distributions for all datasets are shown in \figref{fig:camera_motion_distribution}. It is observed that for a large fraction of the sequences there is no significant camera motion (Cholec80 $76.21\%$, HeiChole $76.2\%$, AutoLaparo $71.29\%$). This finding supports the importance sampling that was introduced in \secref{sec:video_database_and_importance_sampling}. It can further be seen that e.g. left/right and up/down motions are equally distributed.

\subsubsection{Online Camera Motion Estimation}
The results of the online camera motion estimation are summarized in \tabref{tab:estimation_speed}. The deep homography estimation with a Resnet34 backbone executes $11\times$ quicker and has the lowest GPU memory footprint of the GPU accelerated methods. This allows for efficient implementation of the proposed online camera motion estimation in \figref{fig:training_pipeline}.

\begin{table}[htb]
    \caption{Memory footprint and execution time of different camera motion estimators, refer to \secref{sec:online_camera_motion_estimation}.}
    \label{tab:estimation_speed}
    \centering
    \resizebox{\textwidth}{!}{
        \begin{tabular}{lrrr}
            \toprule
            Method & Execution time [s] & Speed-up [a.u.] & Model / Batch [Mb] \\
            \midrule
            Resnet34 & $\mathbf{0.016 \pm 0.048}$ & $\mathbf{11.1}$ & $\mathbf{664} / \mathbf{457}$  \\
            LoFTR \& RANSAC & $0.178 \pm 0.06$ & $1.0$ & $669 / 2412$ \\
            SURF \& RANSAC & $0.131 \pm 0.024$ & $1.4$ & NA \\
            \bottomrule
        \end{tabular}
    }
\end{table}

\subsection{Camera Motion Prediction}
The camera motion prediction results for all datasets are highlighted in \tabref{tab:camera_motion_prediction}. It can be seen that significant improvements over the baseline are achieved on the Cholec80 and HeiChole datasets. Whilst the learned prediction performs better on average than the baseline, no significant improvement is found for the AutoLaparo dataset.

\begin{figure}
    \centering
    \begin{subfigure}[b]{0.49\textwidth}
        \centering
        \includegraphics[width=\textwidth]{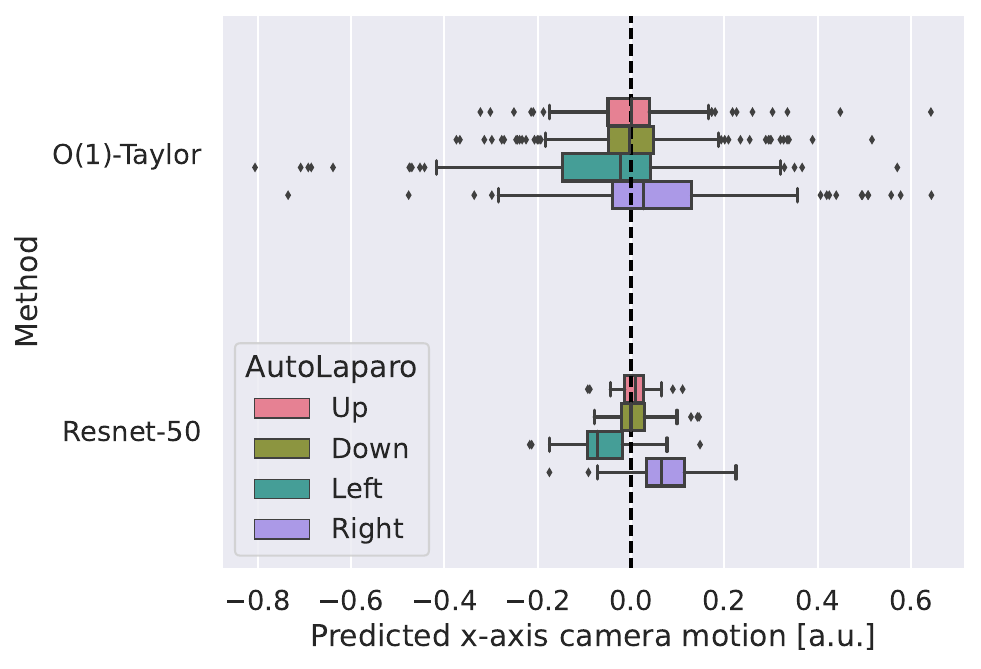}
        \caption{Predicted camera motion along x-axis, scaled by image size to $[-1, 1]$.}
        \label{fig:autolaparo_results_a}
    \end{subfigure}
    \begin{subfigure}[b]{0.49\textwidth}
        \centering
        \includegraphics[width=\textwidth]{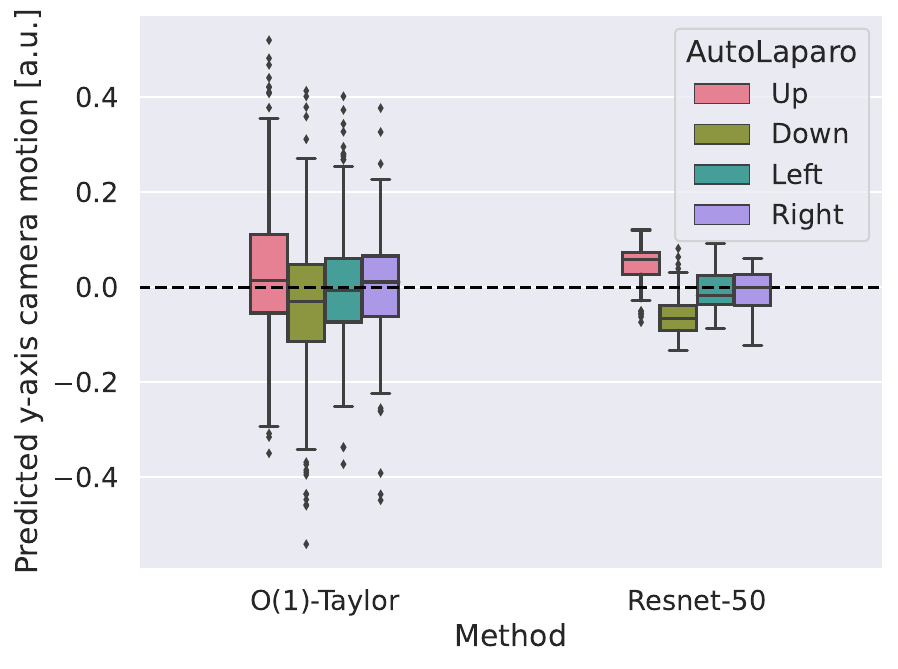}
        \caption{Predicted camera motion along y-axis, scaled by image size to $[-1, 1]$.}
        \label{fig:autolaparo_results_b}
    \end{subfigure}
    \caption{Predicted camera motion on AutoLaparo, refer to \secref{sec:camera_motion_predictor_experiments}. Camera motion predictor trained on Cholec80 with ResNet-50 backbone, see \tabref{tab:camera_motion_prediction}. Shown is the motion of the image center under the predicted homography. Clearly, for videos labeled left/right, the center point is predicted to move left/right and for up/down labels, the predicted left/right motion is centered around zero (a). Same is observed for up/down motion in (b), where left/right motion is zero-centered.}
    \label{fig:autolapato_results}
\end{figure}

The displacement of the image center point under the predicted camera motion for AutoLaparo is plotted against the provided multi-class motion annotations and shown in \figref{fig:autolapato_results}. It can be seen that the camera motion predictions align well with the ground truth labels.

\begin{table}[]
\caption{Camera motion predictor performance, refer to \secref{sec:camera_motion_predictor_experiments}. Taylor baselines predict based on previous estimated motion, ResNets based on images.}
\label{tab:camera_motion_prediction}
\centering
\resizebox{\linewidth}{!}{
\begin{tabular}{|l|l|lllll|}
\hline
\multirow{3}{*}{Dataset} & \multirow{3}{*}{\begin{tabular}[c]{@{}l@{}}Train Size\\ {[}Frames{]}\end{tabular}} & \multicolumn{5}{l|}{Mean Pairwise Distance {[}Pixels{]}}                                                                                                                                          \\ \cline{3-7} 
                         &                                                                                 & \multicolumn{2}{l|}{Taylor}                                                   & \multicolumn{3}{l|}{ResNet (proposed)}                                                                                              \\ \cline{3-7} 
                         &                                                                                 & \multicolumn{1}{l|}{$\mathcal{O}(1)$} & \multicolumn{1}{l|}{$\mathcal{O}(2)$} & \multicolumn{1}{l|}{$18$}                     & \multicolumn{1}{l|}{$34$}                     & $50$                     \\ \hline
Cholec80                 & $3.5e6$                                                                         & \multicolumn{1}{l|}{$27.2 \pm 23.1$}  & \multicolumn{1}{l|}{$36.4 \pm 31.2$}  & \multicolumn{1}{l|}{$\mathbf{14.8} \pm 11.7$} & \multicolumn{1}{l|}{$\mathbf{14.4} \pm 11.4$} & $\mathbf{14.4} \pm 11.4$ \\ \hline
HeiChole                 & $7.6e5$                                                                         & \multicolumn{1}{l|}{$29.7 \pm 26.4$}  & \multicolumn{1}{l|}{$39.8 \pm 35.9$}  & \multicolumn{1}{l|}{$\mathbf{15.8} \pm 12.5$} & \multicolumn{1}{l|}{$\mathbf{15.8} \pm 12.5$} & $\mathbf{15.8} \pm 12.5$ \\ \hline
AutoLaparo               & $5.9e4$                                                                         & \multicolumn{1}{l|}{$19.4 \pm 18.4$}  & \multicolumn{1}{l|}{$25.8 \pm 24.7$}  & \multicolumn{1}{l|}{$\mathbf{11.2} \pm 11.0$} & \multicolumn{1}{l|}{$\mathbf{11.3} \pm 11.0$} & $\mathbf{11.3} \pm 11.0$ \\ \hline
\end{tabular}
}
\end{table}















\begin{figure}
    \centering
    \includegraphics[width=\linewidth]{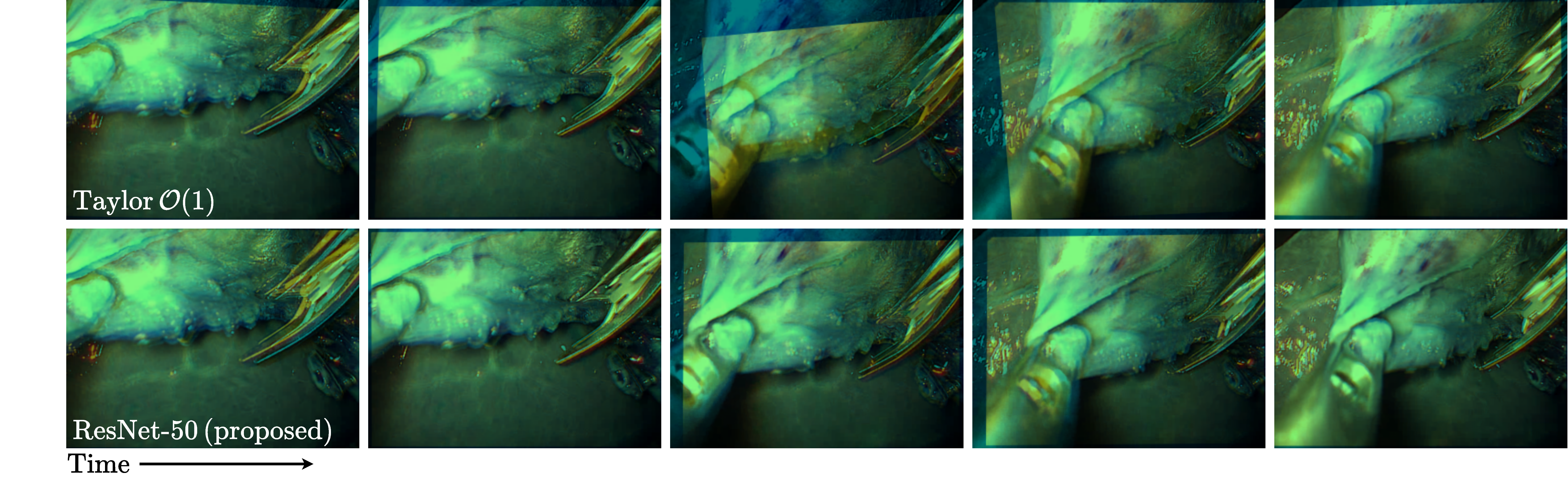}
    \caption{Exemplary camera motion prediction, refer to \secref{sec:camera_motion_predictor_experiments}. In the image sequence, the attention changes from the right to the left tool. We warp the past view (yellow) by the predicted homography and overlay the current view (blue). Good alignment corresponds to good camera motion prediction. Contrary to the baseline, the proposed method predicts the motion well. Data taken from HeiChole test set, ResNet-50 backbone trained on Cholec80, refer \tabref{tab:camera_motion_prediction}.}
    \label{fig:predicted_camera_motion_sequence}
\end{figure}




\section{Conclusion and Outlook}
To the best of our knowledge, this work is the first
to demonstrate that camera motion can indeed be learned from 
retrospective
videos of laparoscopic interventions,
with no manual annotation.
Self-supervision is achieved 
by harvesting image-motion correspondences using a camera motion estimator, see \figref{fig:training_pipeline}. The camera motion predictor is shown to generate statistically significant better predictions over a baseline in \tabref{tab:camera_motion_prediction} as measured using pseudo-ground-truth and on multi-class manually annotated motion labels from AutoLaparo in \figref{fig:autolapato_results}. An exemplary image sequence in \figref{fig:predicted_camera_motion_sequence} demonstrates successful camera motion prediction on HeiChole. These results were achieved through the key finding from \figref{fig:camera_motion_distribution}, which states that most image sequences, i.e. static ones, are irrelevant to learning camera motion. Consequentially, we contribute a novel importance sampling method, as described in \secref{sec:video_database_and_importance_sampling}. Finally, we hope that our open-source commitment will help the community explore this area of research further.

A current limitations of this work is the preview horizon $M$ of length $1$. One might want to extend it for model predictive control. Furthermore, to improve explainability to the surgeon, but also to improve the prediction in general, it would be beneficial to include auxiliary tasks, e.g. tool and organ segmentation, surgical phase recognition, and audio. There also exist limitations for the camera motion estimator. The utilized camera motion estimator is efficient and isolates object motion well from camera motion, but is limited to relatively small camera motions. Improving the camera motion estimator to large camera motions would help increase the preview horizon $M$.

In future work, we will execute this model in a real setup for investigating transferability. This endeavor is backed by \cite{huber2021homography}, which demonstrates how the learned homography could immediately be deployed on a robotic laparoscope holder. It  might proof necessary to fine-tune the presented policy through reinforcement learning with human feedback.


\subsubsection{Acknowledgements}
This work was supported by core and project funding from the Wellcome/EPSRC [WT203148/Z/16/Z; NS/A000049/1; WT101957; NS/A000027/1]. This project has received funding from the European Union's Horizon 2020 research and innovation programme under grant agreement No 101016985 (FAROS project). TV is supported by a Medtronic / RAEng Research Chair [RCSRF1819\textbackslash7\textbackslash34]. SO and TV are co-founders and shareholders of Hypervision Surgical. TV is co-founder and shareholder of Hypervision Surgical. TV holds shares from Mauna Kea Technologies.
%
%
%
\bibliographystyle{splncs04}
\bibliography{literature}

\begin{thebibliography}{10}
\providecommand{\url}[1]{\texttt{#1}}
\providecommand{\urlprefix}{URL }
\providecommand{\doi}[1]{https://doi.org/#1}

\bibitem{agrawal2018automating}
Agrawal, A.S.: Automating endoscopic camera motion for teleoperated minimally
  invasive surgery using inverse reinforcement learning. Ph.D. thesis,
  Worcester Polytechnic Institute (2018)

\bibitem{Budd2022RapidDataset}
Budd, C., Garcia-Peraza~Herrera, L.C., Huber, M., Ourselin, S., Vercauteren,
  T.: {Rapid and robust endoscopic content area estimation: a lean GPU-based
  pipeline and curated benchmark dataset}. Computer Methods in Biomechanics and
  Biomedical Engineering: Imaging and Visualization  (2022).
  \doi{10.1080/21681163.2022.2156393}

\bibitem{cartucho2021visionblender}
Cartucho, J., Tukra, S., Li, Y., S.~Elson, D., Giannarou, S.: Visionblender: a
  tool to efficiently generate computer vision datasets for robotic surgery.
  Computer Methods in Biomechanics and Biomedical Engineering: Imaging \&
  Visualization  \textbf{9}(4),  331--338 (2021)

\bibitem{da2020scan}
Da~Col, T., Mariani, A., Deguet, A., Menciassi, A., Kazanzides, P., De~Momi,
  E.: Scan: System for camera autonomous navigation in robotic-assisted
  surgery. In: 2020 IEEE/RSJ International Conference on Intelligent Robots and
  Systems (IROS). pp. 2996--3002. IEEE (2020)

\bibitem{davenport2019potential}
Davenport, T., Kalakota, R.: The potential for artificial intelligence in
  healthcare. Future healthcare journal  \textbf{6}(2), ~94 (2019)

\bibitem{DeTone2016DeepEstimation}
DeTone, D., Malisiewicz, T., Rabinovich, A.: {Deep Image Homography Estimation}
   (2016), \url{http://arxiv.org/abs/1606.03798}

\bibitem{esteva2019guide}
Esteva, A., Robicquet, A., Ramsundar, B., Kuleshov, V., DePristo, M., Chou, K.,
  Cui, C., Corrado, G., Thrun, S., Dean, J.: A guide to deep learning in
  healthcare. Nature medicine  \textbf{25}(1),  24--29 (2019)

\bibitem{fiorini2022concepts}
Fiorini, P., Goldberg, K.Y., Liu, Y., Taylor, R.H.: Concepts and trends in
  autonomy for robot-assisted surgery. Proceedings of the IEEE
  \textbf{110}(7),  993--1011 (2022)

\bibitem{garcia2022robotic}
Garcia-Peraza-Herrera, L.C., Gruijthuijsen, C., Borghesan, G., Reynaerts, D.,
  Deprest, J., Ourselin, S., Vercauteren, T., Vander~Poorten, E.: Robotic
  endoscope control via autonomous instrument tracking. Frontiers in Robotics
  and AI  (2022)

\bibitem{huber2021homography}
Huber, M., Mitchell, J.B., Henry, R., Ourselin, S., Vercauteren, T., Bergeles,
  C.: Homography-based visual servoing with remote center of motion for
  semi-autonomous robotic endoscope manipulation. In: 2021 International
  Symposium on Medical Robotics (ISMR). pp.~1--7. IEEE (2021)

\bibitem{huber2022deep}
Huber, M., Ourselin, S., Bergeles, C., Vercauteren, T.: Deep homography
  estimation in dynamic surgical scenes for laparoscopic camera motion
  extraction. Computer Methods in Biomechanics and Biomedical Engineering:
  Imaging \& Visualization  \textbf{10}(3),  321--329 (2022)

\bibitem{ji2018learning}
Ji, J.J., Krishnan, S., Patel, V., Fer, D., Goldberg, K.: Learning 2d surgical
  camera motion from demonstrations. In: 2018 IEEE 14th International
  Conference on Automation Science and Engineering (CASE). pp. 35--42. IEEE
  (2018)

\bibitem{kassahun2016surgical}
Kassahun, Y., Yu, B., Tibebu, A.T., Stoyanov, D., Giannarou, S., Metzen, J.H.,
  Vander~Poorten, E.: Surgical robotics beyond enhanced dexterity
  instrumentation: a survey of machine learning techniques and their role in
  intelligent and autonomous surgical actions. International journal of
  computer assisted radiology and surgery  \textbf{11},  553--568 (2016)

\bibitem{kitaguchi2022artificial}
Kitaguchi, D., Takeshita, N., Hasegawa, H., Ito, M.: Artificial
  intelligence-based computer vision in surgery: Recent advances and future
  perspectives. Annals of gastroenterological surgery  \textbf{6}(1),  29--36
  (2022)

\bibitem{li2021data}
Li, B., Lu, B., Lu, Y., Dou, Q., Liu, Y.H.: Data-driven holistic framework for
  automated laparoscope optimal view control with learning-based depth
  perception. In: 2021 IEEE International Conference on Robotics and Automation
  (ICRA). pp. 12366--12372. IEEE (2021)

\bibitem{li2022learning}
Li, B., Lu, B., Wang, Z., Zhong, F., Dou, Q., Liu, Y.H.: Learning laparoscope
  actions via video features for proactive robotic field-of-view control. IEEE
  Robotics and Automation Letters  \textbf{7}(3),  6653--6660 (2022)

\bibitem{li20223d}
Li, B., Wei, R., Xu, J., Lu, B., Yee, C.H., Ng, C.F., Heng, P.A., Dou, Q., Liu,
  Y.H.: 3d perception based imitation learning under limited demonstration for
  laparoscope control in robotic surgery. In: 2022 International Conference on
  Robotics and Automation (ICRA). pp. 7664--7670. IEEE (2022)

\bibitem{loukas2018video}
Loukas, C.: Video content analysis of surgical procedures. Surgical endoscopy
  \textbf{32},  553--568 (2018)

\bibitem{maier2022surgical}
Maier-Hein, L., Eisenmann, M., Sarikaya, D., M{\"a}rz, K., Collins, T.,
  Malpani, A., Fallert, J., Feussner, H., Giannarou, S., Mascagni, P., et~al.:
  Surgical data science--from concepts toward clinical translation. Medical
  image analysis  \textbf{76},  102306 (2022)

\bibitem{marzullo2021towards}
Marzullo, A., Moccia, S., Catellani, M., Calimeri, F., De~Momi, E.: Towards
  realistic laparoscopic image generation using image-domain translation.
  Computer Methods and Programs in Biomedicine  \textbf{200},  105834 (2021)

\bibitem{sandoval2021towards}
Sandoval, J., Laribi, M.A., Faure, J., Breque, C., Richer, J.P., Zeghloul, S.:
  Towards an autonomous robot-assistant for laparoscopy using exteroceptive
  sensors: feasibility study and implementation. IEEE Robotics and Automation
  Letters  \textbf{6}(4),  6473--6480 (2021)

\bibitem{scheikl2023lapgym}
Scheikl, P.M., Gyenes, B., Younis, R., Haas, C., Neumann, G., Wagner, M.,
  Mathis-Ullrich, F.: Lapgym--an open source framework for reinforcement
  learning in robot-assisted laparoscopic surgery. arXiv preprint
  arXiv:2302.09606  (2023)

\bibitem{su2021multicamera}
Su, Y.H., Huang, K., Hannaford, B.: Multicamera 3d viewpoint adjustment for
  robotic surgery via deep reinforcement learning. Journal of Medical Robotics
  Research  \textbf{6}(01n02),  2140003 (2021)

\bibitem{sun2021loftr}
Sun, J., Shen, Z., Wang, Y., Bao, H., Zhou, X.: Loftr: Detector-free local
  feature matching with transformers. In: Proceedings of the IEEE/CVF
  conference on computer vision and pattern recognition. pp. 8922--8931 (2021)

\bibitem{twinanda2016endonet}
Twinanda, A.P., Shehata, S., Mutter, D., Marescaux, J., De~Mathelin, M., Padoy,
  N.: Endonet: a deep architecture for recognition tasks on laparoscopic
  videos. IEEE transactions on medical imaging  \textbf{36}(1),  86--97 (2016)

\bibitem{wagner2021learning}
Wagner, M., Bihlmaier, A., Kenngott, H.G., Mietkowski, P., Scheikl, P.M.,
  Bodenstedt, S., Schiepe-Tiska, A., Vetter, J., Nickel, F., Speidel, S.,
  et~al.: A learning robot for cognitive camera control in minimally invasive
  surgery. Surgical Endoscopy  \textbf{35}(9),  5365--5374 (2021)

\bibitem{wagner2023comparative}
Wagner, M., M{\"u}ller-Stich, B.P., Kisilenko, A., Tran, D., Heger, P.,
  M{\"u}ndermann, L., Lubotsky, D.M., M{\"u}ller, B., Davitashvili, T., Capek,
  M., et~al.: Comparative validation of machine learning algorithms for
  surgical workflow and skill analysis with the heichole benchmark. Medical
  Image Analysis p. 102770 (2023)

\bibitem{wang2022autolaparo}
Wang, Z., Lu, B., Long, Y., Zhong, F., Cheung, T.H., Dou, Q., Liu, Y.:
  Autolaparo: A new dataset of integrated multi-tasks for image-guided surgical
  automation in laparoscopic hysterectomy. In: Medical Image Computing and
  Computer Assisted Intervention--MICCAI 2022: 25th International Conference,
  Singapore, September 18--22, 2022, Proceedings, Part VII. pp. 486--496.
  Springer (2022)

\bibitem{van2018learning}
van Workum, F., Fransen, L., Luyer, M.D., Rosman, C.: Learning curves in
  minimally invasive esophagectomy. World Journal of Gastroenterology
  \textbf{24}(44), ~4974 (2018)

\bibitem{zidane2022robotics}
Zidane, I.F., Khattab, Y., Rezeka, S., El-Habrouk, M.: Robotics in laparoscopic
  surgery-a review. Robotica pp. 1--48 (2022)

\end{thebibliography}

\end{document}